%% file: HIG-14-033_temp.tex
\pdfoutput=1

\documentclass[11pt,twoside,a4paper,cmspaper,final,collab]{cms-tdr}
\def\svnVersion{343674}\def\cmsCernNoTag{CERN-PH-EP/2015-284}\def\cmsCernDate{\today}\def\cmsMessage{Published in Physics Letters B as \href{http://dx.doi.org/10.1016/j.physletb.2016.05.003}{\doi{10.1016/j.physletb.2016.05.003.}}}

\begin{document}\cmsNoteHeader{HIG-14-033}

\hyphenation{had-ron-i-za-tion}
\hyphenation{cal-or-i-me-ter}
\hyphenation{de-vices}

\RCS$Revision: 343492 $
\RCS$HeadURL: svn+ssh://svn.cern.ch/reps/tdr2/papers/HIG-14-033/trunk/HIG-14-033.tex $
\RCS$Id: HIG-14-033.tex 343492 2016-05-19 20:06:10Z abdollah $
\newlength\cmsFigWidth
\ifthenelse{\boolean{cms@external}}{\setlength\cmsFigWidth{0.85\columnwidth}}{\setlength\cmsFigWidth{0.4\textwidth}}
\ifthenelse{\boolean{cms@external}}{\providecommand{\cmsLeft}{top\xspace}}{\providecommand{\cmsLeft}{left\xspace}}
\ifthenelse{\boolean{cms@external}}{\providecommand{\cmsRight}{bottom\xspace}}{\providecommand{\cmsRight}{right\xspace}}
\ifthenelse{\boolean{cms@external}}{\providecommand{\breakhere}{\relax}}{\providecommand{\breakhere}{\linebreak[4]}}
\newcommand{\tauh}{\ensuremath{\Pgt_\mathrm{h}}\xspace}
\newcommand{\taumu}{\ensuremath{\Pgt_{\Pgm}}\xspace}
\newcommand{\taue}{\ensuremath{\Pgt_{\Pe}}\xspace}

\newcommand{\mtau}{\ensuremath{\Pgm\Pgt_\Ph}\xspace}
\newcommand{\etau}{\ensuremath{\Pe\Pgt_\Ph}\xspace}
\newcommand{\emu}{\ensuremath{\Pe\Pgm}\xspace}
\newcommand{\ee}{\ensuremath{\Pe\Pe}\xspace}
\newcommand{\mumu}{\ensuremath{\Pgm\Pgm}\xspace}
\newcommand{\tautau}{\ensuremath{\Pgt_\Ph\Pgt_\Ph}\xspace}

\newcommand{\MT}{\ensuremath{M_\mathrm{T}}\xspace}
\providecommand{\mH}{\ensuremath{m_{\PH}\xspace}}

\providecommand{\NA}{\text{---}\xspace}

\renewcommand{\ptvecmiss}{\ensuremath{\vec{p}_{\mathrm{T}}^{\kern2pt\text{miss}} }\xspace}
\newcommand{\ptvecell}{\ensuremath{\vec{p}_{\mathrm{T}}^{\kern2pt\ell} }\xspace}
\newcommand{\ptvece}{\ensuremath{\vec{p}_{\mathrm{T}}^{\kern2pt\mathrm{e}} }\xspace}
\newcommand{\ptvecmu}{\ensuremath{\vec{p}_{\mathrm{T}}^{\kern2pt\mu} }\xspace}

\cmsNoteHeader{HIG-14-033}
\title{Search for a low-mass pseudoscalar Higgs boson produced in association with a \texorpdfstring{\bbbar}{b-bbar} pair in pp collisions at \texorpdfstring{$\sqrt{s} = 8\TeV$}{sqrt(s) = 8 TeV}}

\date{\today}

\abstract{A search is reported for a light pseudoscalar Higgs boson decaying to a pair of $\tau$ leptons, produced in association with a  \bbbar pair, in the context of two-Higgs-doublet models. The results are based on pp collision data at a centre-of-mass energy of 8\TeV collected by the CMS experiment at the LHC and corresponding to an integrated luminosity of 19.7\fbinv. Pseudoscalar boson masses between 25 and 80\GeV are probed.  No evidence for a pseudoscalar boson is found and upper limits are set on the product of cross section and  branching fraction to $\tau$ pairs between 7 and 39\unit{pb} at the 95\% confidence level.
This excludes pseudoscalar A bosons with masses between 25 and 80\GeV, with SM-like Higgs boson negative couplings to down-type fermions, produced in association with \bbbar pairs, in Type II, two-Higgs-doublet models.}

\hypersetup{%
pdfauthor={CMS Collaboration},%
pdftitle={Search for a low-mass pseudoscalar Higgs boson produced in association with a b-bbar pair in pp collisions at sqrt(s) = 8 TeV},%
pdfsubject={CMS},%
pdfkeywords={CMS, physics, Higgs, }}

\maketitle

\section{Introduction}
\label{Introduction}

The discovery of a new boson with a mass close to 125\GeV~\cite{ATLAS-HIGGS-DISCOVERY,CMS-HIGGS-DISCOVERY,CMS-HIGGS-LongPaper}, consistent with the standard model (SM) Higgs boson, has shed light on one of the most important questions of physics:
the origin of the mass of elementary particles.
Although all the measurements made up to now are in impressive agreement with the predictions of the SM~\cite{ATLAS-MASS,HIG-14-009},
the SM cannot address several crucial issues such as the hierarchy problem, the origin of the matter-antimatter asymmetry and the nature of dark matter~\cite{Wess:1974tw,Cheng:2003ju,Appelquist:2000nn,DarkMatt}.
Theories predicting new physics beyond the standard model have been proposed to address these open questions.
Many of them  predict the existence of more than one Higgs boson.

Two-Higgs-doublet models (2HDM)~\cite{2HDM-Lee,2HDM-Deshpande,2HDM-Haber,Gunion:1989we,Branco:2011iw} are a particularly simple extension of the SM.  Starting with the two doublet fields $\Phi_1$ and $\Phi_2$ and assuming an absence of CP violation in the Higgs sector, after SU(2)$_{L}$ symmetry breaking five physical states are left: two CP-even (h and H), one CP-odd (A), and two charged ($\mathrm{H^{\pm}}$) bosons.  To avoid tree-level flavour changing neutral currents, one imposes a Z$_{2}$ symmetry according to which the Lagrangian is required to be invariant under $\Phi_1\to \Phi_1$,  $\Phi_2\to -\Phi_2$. The result is four distinct classes of models, corresponding to different patterns of quark and lepton couplings. The most commonly considered are Type~I and Type~II. In Type~I, all quarks and leptons obtain masses from $\langle\Phi_1\rangle$.  In Type~II, up-type quarks masses are derived from $\langle\Phi_1\rangle\equiv v_1$ and down-type quarks and charged leptons masses are derived from $\langle\Phi_2\rangle\equiv v_2$.
In the limit of an exact Z$_{2}$ symmetry~\cite{Z2Sym}, the Higgs sector of a 2HDM can be described by six parameters: four Higgs boson masses ($m_{\Ph}$, $m_{\PH}$, $m_{\PSA}$, and $m_{\PH^{\pm}}$), the ratio of the vacuum expectation values of the two doublets ($\tan\beta \equiv v_2/v_1 $) and the mixing angle $\alpha$  of the two neutral CP-even Higgs states. Allowing a soft breaking of the Z$_{2}$ symmetry introduces a new Higgs mixing parameter ${m_{12}^{2}}$~\cite{2HDM-Deshpande}. In the ``decoupling limit" of 2HDMs  ~\cite{2HDM-Haber-Nir,Gunion-decoupling}, the masses $m_{\PH}$, $m_{\PSA}$, and $m_{\PH^{\pm}}$ are all large, $\cos(\beta - \alpha) \ll 1$, and $\Ph$ is the observed boson at 125\GeV and is SM-like. An SM-like $\Ph$ or $\PH$ at 125\GeV can also be obtained in the ``alignment limit"~\cite{2HDM-Haber-Nir,Gunion-decoupling} without the other bosons being heavy.  This is an interesting case and can be compatible with the SM-like Higgs boson total width measurements and branching fractions even if one or more of the light Higgs bosons have a mass below half of 125\GeV provided one adjusts the model parameters so that the branching fraction of the SM Higgs boson to pairs of light Higgs bosons is very small.  This scenario can be tested at the CERN LHC by searching for singly produced light bosons decaying to a pair of $\tau$ leptons with large cross sections. In Type~II 2HDMs, if the Higgs coupling to the third generation of quarks is enhanced, as happens at large $\tan\beta$, a large production cross section is expected for the production of the low-mass $\PSA$ boson in association with \bbbar. The cross section is of the order of 1\unit{pb} for regions of the 2HDM parameter space with $\sin(\beta - \alpha) \approx 1$, $\cos(\beta - \alpha) > 0$  and small $m_{12}^{2}$.
The cross section can be much larger, between 10 and 100\unit{pb}, for some other regions of the parameter space,  i.e.~$\sin(\beta \pm \alpha) \approx 1$, $\cos(\beta - \alpha) < 0$ and $\tan\beta > 5$ ~\cite{JGunion13TeV,JGunion8TeV}, where  the coupling of the SM-like h boson to down-type fermions is negative (``wrong sign" Yukawa coupling). Consequently, given the large production cross section of the $\PSA$ boson in such scenarios, the LHC data are sensitive to its presence for some combinations of model parameters.

Previous searches for di-$\tau$ resonances~\cite{Aad:2014vgg,Khachatryan:2014wca} have mainly focused on masses greater than the mass of the $\PZ$ boson, for example in the context of the minimal supersymmetric standard model (MSSM)~\cite{Fayet:1974pd,Fayet:1976et,Fayet:1977yc}, which is a highly constrained  2HDM of Type~II.  In fact, a light pseudoscalar Higgs boson is excluded in the MSSM, but an $\PSA$ boson can still have quite a low mass in general 2HDMs, even given all the constraints from LEP, Tevatron and LHC data~\cite{JGunion13TeV,JGunion8TeV}.

This letter presents a search for a low-mass pseudoscalar Higgs boson produced in association with a \bbbar pair and decaying to a pair of $\tau$ leptons. Associated production of the $\PSA$ boson with a \bbbar pair has the advantage that there is a higher signal over background ratio relative to gluon-gluon fusion production. Such a signature is also relevant in the context of light pseudoscalar mediators and coy dark sectors~\cite{Kozaczuk:2015bea}.
 The analysis is based on pp collision data at a centre-of-mass energy of 8\TeV recorded by the CMS experiment at the LHC in 2012. The integrated luminosity amounts to 19.7\fbinv.
The $\tau$ leptons are reconstructed via their muon, electron and hadronic decays. In the following, the terms leptons refer to electrons and muons, whereas $\tau$s that decay into hadrons+$\nu_\tau$ are denoted by $\tauh$. The invariant mass distributions of the $\tau$ pairs in all three channels are used to search for pseudoscalar bosons with masses between 25 and 80\GeV.

\section{The CMS detector and event samples}
\label{CMS-Detector}

The central feature of the CMS apparatus is a superconducting solenoid of 6\unit{m} internal diameter, providing a
magnetic field of 3.8\unit{T}. Within the solenoid volume are a silicon pixel and strip tracker,
a lead tungstate crystal electromagnetic calorimeter (ECAL), and a brass and scintillator hadron calorimeter (HCAL),
each composed of a barrel and two endcap sections. Muons are detected in gas-ionisation detectors embedded in the steel
flux-return yoke outside the solenoid. Extensive forward calorimetry complements the coverage provided by the barrel and endcap detectors.
A detailed description of the CMS detector, together with a definition of the coordinate system used and the relevant kinematic variables, can be found in Ref.~\cite{Chatrchyan:2008zzk}.

The first level of the CMS triggering system (Level-1), composed of custom hardware processors, uses
information from the calorimeters and the muons detectors to select the most interesting events in a fixed time interval of less than 4\mus. The high-level trigger (HLT) processor farm further decreases the event rate from around 100\unit{kHz} to less than 1\unit{kHz}, before data storage.

A set of Monte Carlo (MC) simulated events is used to model the signal and backgrounds.
Drell--Yan, W boson production
associated to additional jets, production of top quark pairs (\ttbar), and diboson ($\PW\PW$, $\PW\PZ$~and $\PZ\PZ$) backgrounds are  generated using the leading order (LO) \textsc{MadGraph} 5.1 package~\cite{MadGraph}.
Single top quark samples are produced using the next-to-leading-order (NLO) generator \POWHEG (v1.0)~\cite{POWHEG}.
Simulated samples of gluon-gluon fusion to $\bbbar\PSA$ signal events are generated with  \PYTHIA6.426~\cite{pythia6_4}
for masses between 25 and 80\GeV in 5\GeV steps. As no loop is involved at leading order in the $\bbbar\PSA$ production process, the product of acceptance and efficiency for signal only depends on the A boson mass, with no dependence on other model parameters.
The simulated samples are produced using the CTEQ6L1 parton distribution function (PDF) set~\cite{CTEQ}.
All the generated signal and background samples are processed with the simulation of the CMS detector based on \GEANT4~\cite{geant}.

Additional events are added to the MC-simulated events, with weights corresponding to the luminosity profile in data, to simulate LHC conditions and the presence of other soft pp interactions (pileup) in the same or neighbouring bunch crossings of the main interaction. Finally, identical algorithms and procedures are used to reconstruct both simulated events and the collected data.

\section{Event reconstruction}
\label{sec:Event-Reco}

Event reconstruction is based on the particle-flow (PF) algorithm~\cite{PFT-09-001,PFT-10-001}, which aims to exploit the information from all subdetectors
 to identify individual particles (PF candidates): charged and neutral hadrons, muons, electrons, and photons.
Complex objects, such as $\tau$ leptons that decay into hadrons and a neutrino, jets, and the imbalance in the transverse momentum in the event
are reconstructed from PF candidates.

The deterministic annealing algorithm ~\cite{DAA1,DAA2} is used to reconstruct the collision vertices. The vertex with the maximum sum of squared transverse momenta ($\pt^2$) of all associated tracks is considered as the primary vertex.
Muons, electrons, and $\tauh$s are required to originate from the primary collision vertex.

Muon reconstruction starts by matching  tracks in the silicon tracker
with tracks in the outer muon spectrometer~\cite{Chatrchyan:2012xi}.
A global muon track is fitted to the hits from both tracks.
A preselection is applied to these muon tracks that includes requirements on their impact parameters, to distinguish genuine prompt muons from spurious
muons or muons from cosmic rays. In addition, muons are required to pass isolation criteria to separate  prompt muons from those associated with a jet, usually from the semi-leptonic decays of heavy quarks.
The muon relative isolation is defined as the following~\cite{Chatrchyan:2008zzk}:

\begin{equation}
\ifthenelse{\boolean{cms@external}}
{
\begin{split}
& I_{\text{rel}} = \left[ \sum\limits_{\mathrm{charged}} \pt \right. + \\ & \left. \max \left(0, \sum\limits_{\mathrm{neutral}} \pt + \sum\limits_{\gamma} \pt -\frac{1}{2} \sum\limits_{\mathrm{charged,PU}} \pt\right)  \right] / p_{\kern-1pt\mathrm{T}}^\mu,
\end{split}
}
{
I_{\text{rel}}=\left[ \sum\limits_{\mathrm{charged}} \pt + \max \left(0, \sum\limits_{\mathrm{neutral}} \pt + \sum\limits_{\gamma} \pt -\frac{1}{2} \sum\limits_{\mathrm{charged,PU}} \pt\right)\right] / p_{\kern-1pt\mathrm{T}}^\mu,
}
\label{eq:Iso}
\end{equation}

where all sums are over the scalar \pt of particles inside a cone with size of
$\Delta R= \breakhere\sqrt{\smash[b]{(\Delta\eta)^2+(\Delta\phi)^2}}=0.4$ relative to the muon direction, where $\eta$ is the pseudorapidity and $\phi$ is the azimuthal angle (in radians) in the plane transverse to the beam axis, and ``charged" corresponds to charged hadrons, muons, and electrons originating from the primary vertex,  ``neutral" refers to neutral hadrons and ``charged,PU" refers to charged hadrons, muons, and electrons originating from other reconstructed vertices.
 The last of these sums is used to subtract the neutral pileup component in the computation, and the factor of 1/2 reflects the approximate ratio of neutral to charged particles in jets~\cite{Chatrchyan:2011ds}.

Electron reconstruction starts from ECAL superclusters, which are groups of one or more associated clusters of energy deposited in the ECAL.
Superclusters are matched to track seeds in the inner tracker (the closest layers of the tracker to the interaction point) and electron tracks are formed from those.
Trajectories are reconstructed based on the modelling of electron energy loss due to bremsstrahlung,
and are fitted using the Gaussian sum filter algorithm~\cite{Adam2005}.
Electron identification is based on a multivariate (MVA) boosted decision tree technique~\cite{Hocker:2007ht} to discriminate genuine electrons
from jets misidentified as electrons~\cite{EleMVA}. The most powerful variables for the discrimination of $\tauh$ candidates are the
ratio of energy depositions in the ECAL and HCAL, the angular difference between the track and supercluster, and the distribution of energy depositions in the electron shower.
Relative isolation is defined in an analogous way to that of Eq. (\ref{eq:Iso}) and is used to distinguish prompt electrons from electrons within a jet.

Jets are reconstructed from PF candidates using the anti-$\kt$~\cite{AntiKT} algorithm with a distance parameter of 0.5, in the {\sc FastJet} package~\cite{FASTJet}.
Several corrections are applied to the jet energies to reduce the effect of pileup  and correct for the nonlinear response of the calorimeters~\cite{Chatrchyan:2011ds}.
To identify and reject jets from pileup, an MVA discriminator is defined based on information from the vertex and the jet distribution~\cite{JME-13-005}.
Jets identified as originating from a b quark, called b-tagged jets, are identified using the combined secondary vertex (CSV) algorithm~\cite{CSV},
which is based on a likelihood technique, and
exploits information such as the impact parameters of charged-particle tracks and the properties of reconstructed decay vertices.

The hadron-plus-strips (HPS) algorithm~\cite{TAU-11-001,Khachatryan:2015dfa} is used to reconstruct the $\tauh$ candidates.
It starts from a jet,
and
searches for candidates produced by the main hadronic decay modes of the $\tau$ lepton: either directly to one charged hadron, or via intermediate $\rho$ and $\mathrm{a_1(1280)}$ mesons to one
charged hadron plus one or two neutral pions, or three charged hadrons with up to one neutral pion.
The charged hadrons are usually long-lived pions, while the neutral pions decay rapidly into two photons.
The HPS algorithm takes into account the possible conversion of photons into $\Pe^+\Pe^-$ pairs in material in front of the ECAL, and their corresponding bremsstrahlung in the magnetic field with consequent broadening of the distribution of the shower.
Strips are formed from energy depositions in the ECAL arising from electrons and photons. The strip sizes in ECAL  are $0.05{\times}0.20$ in $\eta \times \phi$.
The $\tauh$ decay modes are reconstructed by combining the charged hadrons with ECAL strips.
Neutrinos produced in $\tauh$ decays are not reconstructed but contribute to \MET.
Isolation requirements based on an MVA technique take into account the \pt of PF candidates around the $\tau$ lepton direction and information related to its lifetime, such as the transverse impact parameter of the leading track of the $\tauh$ candidate and its significance for decays to one charged hadron
or the distance between the $\tauh$ production and decay vertices and its significance for decays to three charged hadrons.
Electrons can be misidentified as $\tauh$ candidates with one track and ECAL strip.
An MVA discriminator based on properties of the reconstructed electron, such as the distribution of the shower and the ratio of the ECAL and HCAL deposited energies, is used to
improve pion/electron separation.
Finally, another MVA discriminator is used to suppress muons reconstructed as $\tauh$ candidates with one track. It exploits information about
the energy deposited in the calorimeters with $\tauh$ candidates, as well as hits and segments reconstructed in the muon spectrometers that can be
matched to the components of the $\tauh$.

The missing transverse momentum vector \ptvecmiss is defined as the projection on the plane perpendicular to the beams of the negative vector sum of the momenta of all reconstructed particles in an event. Its magnitude is referred to as \ETmiss.
To improve the resolution, and reduce the effect of pileup, a \ptvecmiss based on an MVA  regression technique~\cite{CMS_AN_2012-226} is used, which takes into account several collections of particles from different vertices.

The invariant mass of the $\tau$ pair ($m_{\tau\tau}$) is used as the observable for the statistical interpretation of results in all channels and is reconstructed using the \textsc{SVFit} algorithm~\cite{SVFIT-New}.
The {\sc SVFit} algorithm uses a maximum likelihood technique where the likelihood takes as input the four-momenta of the visible decay products of the $\tau$,
the projection of \ptvecmiss along the $x$- and $y$-axes, as well as
the covariance matrix of the components of \ptvecmiss.

The relative $m_{\tau\tau}$ resolution obtained through the \textsc{SVFit} algorithm is about 15\%  over the whole mass range. It is slightly higher for the \emu~channel because of the presence of one additional neutrino.

\section{Event selection}
\label{sec:Event-Select}

Three di-$\tau$ final states are considered: \mtau, \etau, and \emu. The \mumu~and \ee~final states are discarded because of their small branching
fractions and large backgrounds, while \tautau~is not considered because of inefficiencies due to the trigger
 threshold.

The selection of events in the \mtau~or \etau~final state starts from
a trigger that requires a combination of a muon or electron with $\pt > 17$  or 22\GeV, respectively, and an isolated $\tauh$ with  $\pt > 20\GeV$. This combined trigger is seeded by a single muon or electron, with  $\pt >16$  or   20\GeV at Level-1.
The offline selection requires a muon or electron with $\pt >  18$ or $ 24\GeV$, respectively, and $\abs{\eta} < 2.1$, and an oppositely charged $\tauh$ candidate
with $\pt >22\GeV$ and $\abs{\eta} < 2.3$.
Leptons are required to pass a tight identification~\cite{Chatrchyan:2012xi,EleMVA} and have a relative isolation, $I_{\text{rel}}$, ${<}0.1$.
The $\tauh$ candidates have to pass a tight
working point of the MVA discriminant that combines isolation and lifetime information (resulting in a $\tauh$ reconstruction and isolation efficiency of about 30\% and a jet to $\tauh$ misidentification rate between 0.5 and 1.0 per mille), as well as the requirements to suppress electron and muon candidates misidentified as $\tauh$, described in Section~\ref{sec:Event-Reco}.
Leptons and $\tauh$ candidates are required to be separated by $\Delta R > 0.5$.
Events with additional identified and isolated electrons or muons are discarded.
To suppress $\PW$+jets and \ttbar backgrounds,  the transverse mass between
 the lepton transverse momentum \ptvecell  and \ptvecmiss, defined in Eq.~\ref{eq:TMass}, is required to be smaller than 30\GeV,
\begin{equation}
\MT(\ell,\ptvecmiss) = \sqrt{2\pt^{\ell}\ETmiss (1- \cos\Delta\phi) },
\label{eq:TMass}
\end{equation}
where
$\Delta\phi$ is the azimuthal angle between the lepton transverse momentum and the \ptvecmiss vectors.

Events selected in the \emu~channel must pass a trigger that requires a combination of an electron and a muon, with $\pt > 17 (8)\GeV$ for the leading (subleading) lepton.
Depending on the flavour of the leading lepton that passes the trigger selection, events are required to have either a muon with $\pt > 18\GeV$ and an electron with $\pt > 10\GeV$, or a muon with $\pt >10\GeV$ and an electron with $\pt > 20\GeV$.
The fiducial regions for muons (electrons) are defined by $\abs{\eta} < 2.1(2.3)$.
Additionally, leptons with opposite charge are selected and required to be spatially separated by $\Delta R > 0.5$.

The muons and electrons are required to be isolated, with relative isolation less than $0.15$ in the barrel
($\abs{\eta}<1.479$) and less than $0.1$ in the endcaps  ($\abs{\eta}>1.479$).
In addition, both muons and electrons are required to pass the tight identification criteria as described in Section~\ref{sec:Event-Reco}.
Events having additional identified and isolated leptons are vetoed, similarly to the \mtau~and \etau~channels.
To reduce the large \ttbar background in the \emu~final state, a linear combination of the $P_{\zeta}$ and $P_{\zeta}^{\text{vis}}$ variables~\cite{CDFrefPzeta} is used.
 $P_{\zeta}$ and $P_{\zeta}^{\text{vis}}$ are defined as follows:
 
\begin{equation}
\ifthenelse{\boolean{cms@external}}
{
\begin{split}
P_{\zeta}  = \left(\ptvecmu + \ptvece + \ptvecmiss \right) \cdot \hat{\zeta}
& \quad \text{and} \\
P_{\zeta}^{\text{vis}}  = \left( \ptvecmu + \ptvece \right) \cdot \hat{\zeta} &,
\end{split}
}
{
P_{\zeta} = \left( \ptvecmu + \ptvece + \ptvecmiss \right) \cdot \hat{\zeta}
\qquad \text{and} \qquad
P_{\zeta}^{\text{vis}} = \left( \ptvecmu + \ptvece \right) \cdot \hat{\zeta},
}
\label{eq:PzetaDefinition}
\end{equation}

where $\hat{\zeta}$ is the unit vector of the axis bisecting the angle between \ptvecmu and \ptvece
of the muon and electron candidates, respectively.  These variables take into account the fact that
the neutrinos produced in $\tau$ decays are mostly collinear with the visible $\tau$  decay products, but this is not true for
neutrinos from the other sources, nor for misidentified  $\tauh$ candidates from background.
The linear combination $P_{\zeta} - \alpha P_{\zeta}^{\text{vis}}$ is required to be greater than $- 40\GeV$, with an optimal value of  $\alpha$ of 1.85, determined in the CMS search for a MSSM Higgs boson in the $\tau\tau$ final state~\cite{Khachatryan:2014wca}.
To further reduce \ttbar and electroweak backgrounds in the \emu~final state, the $\MT$ between the dilepton transverse momentum and \ptvecmiss, defined as in Eq.~\ref{eq:TMass}, is required to be less than 25\GeV.

In addition to the above selections, events in all channels are also required to have at least one b-tagged jet with $\pt >20\GeV$ and $\abs{\eta} < 2.4$, which passes the working point of the CSV {b-tagging} discriminant (corresponding to b-tagging efficiency of about 65\% and light-jet misidentification rate of about 1\%) and the pileup MVA discriminant for jets, and is separated by at least $\Delta R=0.5$ from the signal leptons.

\section{Background estimation}
\label{sec:BGEstimation}

One of the main backgrounds in all three channels is $Z/\gamma^*\to \tau\tau$.
Drell--Yan events with invariant mass larger than 50\GeV are modelled using ``embedded" event samples, as follows:
$\PZ\to\mu\mu$ events are selected in data  with an invariant mass larger than 50 GeV to remove the mass range biased by a trigger requirement. The reconstructed muons are replaced by simulated $\tau$ leptons that are subsequently decayed via {\TAUOLA}~\cite{tauola}.
To model the detector response to the $\tau$ decay products the GEANT based detector simulation is used.
Jets, \ptvecmiss, and $\tauh$ are then reconstructed, while lepton isolations are recomputed~\cite{HIG-13-004}.
This substantially reduces the uncertainties related to the modelling of the \MET, the jet energy scale, and the b jet efficiency.
Low-mass $\PZ/\gamma^*\to \tau\tau$ events, which cannot be covered by the embedded samples,  are taken directly from a simulated sample.

Multijet events originated by QCD processes comprise another major background, especially at low di-$\tau$ mass. The contribution of the QCD multijet
background arises from $\text{jet}\to\tauh$ misidentification and to a lesser extent from $\text{jet}\to \mu$ and $\text{jet}\to \Pe$ misidentification,
depending on the final state.
Other contributions are due to the presence of muons or electrons from the semi-leptonic decays of heavy flavour quarks.
This background is estimated from data.

Multijet background normalisation in the \mtau~and \etau~final states is determined from a sample defined in the same way as the signal selection described in Section~\ref{sec:Event-Select}, except that
the lepton and the $\tauh$ candidate are required to have electric charge of same sign (SS). The events with the SS selection are dominated by multijets,
and the limited contribution from the other processes is subtracted using predictions from simulated events.
To take into account the difference in the multijet normalisation between the SS and opposite-sign (OS) regions, an OS/SS extrapolation factor is used to multiply the multijet yield in the SS region. This factor is measured
in signal-free events selected with inverted lepton isolations ($0.2<I_{\text{rel}}<0.5$) and a relaxed $\tauh$ isolation.
The OS/SS extrapolation factor is parameterised as a function of $m_{\tau\tau}$, and fitted with an exponentially decreasing function.
This ratio is approximately equal to 1.2 for di-$\tau$ masses of 20\GeV, and decreases to about 1.1 for masses above 50\GeV.

The $m_{\tau\tau}$ distribution for the QCD multijet background is obtained from a control region in data by inverting the lepton isolation and relaxing the $\tauh$ isolation.
These two selections are required to attain a control region populated with QCD multijet events and obtain a sufficiently smooth $m_{\tau\tau}$ distribution.
A correction has been applied to account for the differences between the nominal selection and the selection used to estimate the QCD multijet $m_{\tau\tau}$ distribution.
The correction depends on the $\tauh$ misidentification rate (the probability for a $\tauh$,  that passes a looser isolation requirement, to pass  the tight isolation selection). This rate is parameterised as a function of the \pt of the $\tauh$ in three bins of pseudorapidity. It was checked that the $m_{\tau\tau}$ distributions obtained when the lepton isolation is inverted and the $\tauh$ isolation is relaxed, are consistent within statistical uncertainties with the normal search procedure.

In the \emu~final state, the QCD multijet background is measured simultaneously with other backgrounds using misidentified leptons in data, through a ``misidentified-lepton" method~\cite{HIG-13-004}, and requiring at least one jet misidentified as a lepton.
The probability for loosely preselected leptons, mainly dominated by leptons within jets, to be identified as good leptons is measured in samples depleted of isolated leptons
as a function of the \pt and $\eta$.
Weights obtained from this measurement are applied to events in data
with electrons and muons passing the loose preselection but not the nominal selection criteria, to extract the QCD multijet
background contribution.

In the \mtau~and \etau~final states, the $\PW$+jets background arises from  events with a  genuine isolated and identified lepton
from the leptonic decay of a $\PW$  boson and a jet misidentified as a $\tauh$.
Its contribution is highly suppressed by requiring the $\MT$ of the lepton and \ptvecmiss of Eq. (\ref{eq:TMass}) to be ${<}30\GeV$ (low-$\MT$ region).
The $\PW$+jets normalisation is determined from collision data using the yield in the high-$\MT$ (${>}70\GeV$) sideband, multiplied by an extrapolation factor that
is the ratio of the $\PW$+jets events in the  high- and low-$\MT$ regions in simulated events.
The small contribution from other backgrounds in events selected with high-$\MT$ selection is subtracted using the prediction from simulations.
The distribution of $m_{\tau\tau}$ for the $\PW$+jets background is taken from simulation. A correction to the distribution, measured in a sample enriched in $\PW$+jets and
as a function of the $\pt$ of the lepton originating from the $\PW$ boson, is applied
to correct the differences between observed and simulated events.
In the \emu~final state, the $\PW$+jets background is estimated together with the backgrounds that contain at least one jet misidentified as a lepton, such as QCD multijets,
as previously described.

The $\PZ/\gamma^*\to\mu\mu $ and $\PZ/\gamma^*\to\Pe\Pe$  processes  contribute, respectively, to the \mtau~and \etau~final states,  because of the misidentification of a lepton as a $\tauh$.
The normalisation and the distribution of $m_{\tau\tau}$ for these backgrounds are obtained from simulation.

The presence of genuine b jets from top quark decays makes the \ttbar background contribution important. The \ttbar background has true $\tauh$ $\approx$70\% of the times and misidentified $\tauh$ in $\approx$30\% of the times.
The distribution of $m_{\tau\tau}$ for \ttbar events is taken from simulation, but normalised to the measurement of the \ttbar cross section~\cite{Khachatryan:2015oqa}. A reweighting is applied to generated \ttbar events to improve the modelling of the top quark $\pt$ spectrum. This reweighting only depends on the simulated $\pt$ of top and antitop quarks~\cite{Khachatryan:2015oqa}, and has a negligible impact on the final results. In addition, the $m_{\tau\tau}$ distributions observed in data and predicted by MC simulations are compared in a region with high purity of \ttbar events, and depleted in signal, obtained by raising the \pt threshold of the leptons and $\tauh$, and requiring at least two b-tagged jets with a higher \pt threshold than that used in event selections described in Section~\ref{sec:Event-Select}. Good agreement is found between distributions in data and MC simulation.

Single top quark, diboson ($\PW\PW$, $\PW\PZ$, $\PZ\PZ$), and SM Higgs backgrounds represent a small fraction of the total background, and are taken from simulations and normalised to the NLO cross sections~\cite{Chatrchyan:2014tua,MCFMdiBosonXsection,HIG-13-004}.

Scale factors to correct for residual discrepancies between data and MC simulation related to the lepton triggering, identification, and isolation are applied to the signal and the backgrounds estimated from MC simulations. These correction factors are determined using the ``tag-and-probe" technique~\cite{tag-and-probe,TAU-11-001,Khachatryan:2015dfa},  which relies on the presence of two leptons from $\PZ$ boson decays.  No correction factor is applied to the $\tauh$ candidate nor to the selected b jet, as the corrections are found to be consistent with unity. The uncertainties related to these scale factors are described in Section~\ref{sec:Systematics}.

\section{Systematic uncertainties}
\label{sec:Systematics}

The results of the analysis are extracted from a fit based on the $m_{\tau\tau}$ distributions in each final state, as discussed in Section~\ref{sec:Results}.
Systematic uncertainties in the fit affect the normalisation
or the shape of the $m_{\tau\tau}$ distribution for the signal and backgrounds. The normalisation uncertainties are summarised in Table~\ref{table:uncertainties}.

The uncertainty in normalisation that affects the signal and most of the simulated backgrounds is related to the integrated luminosity at 8\TeV,
which is measured with a precision of 2.6\%~\cite{LUM-13-001}.
Uncertainties in muon and electron identification and trigger efficiency, as well as in the $\tauh$ identification efficiency, are determined using the ``tag-and-probe" technique~\cite{tag-and-probe,TAU-11-001,Khachatryan:2015dfa}.  These uncertainties are about 2\% for muon and electron and 8\% for $\tauh$.
Changes in acceptance due to the uncertainty in the b tagging efficiency and the b mistag rate range from 1 to 9\% depending on the process.
To estimate the uncertainty in  the $\PW$+jets normalisation, the uncertainty in the extrapolation factor from the high-\MT~sideband to the signal region
is obtained by varying \MET and its resolution
by their uncertainties, leading to a 30\% uncertainty.
The uncertainty in the normalisation of QCD multijet background is obtained by adding the
statistical uncertainty related to the sample size of the QCD multijet-dominated control region
in quadrature with the uncertainty in the extrapolation factor from the control region to the signal region; this amounts to 20\%.
The normalisation uncertainty for the \ttbar~background amounts to 10\%; it is determined from a control region where both W bosons originating from the top and antitop quarks decay to $\tau$ leptons\cite{HIG-13-004}. Uncertainties related to the diboson background cross section amount to  15\% ~\cite{Chatrchyan:2013oev}.

A 30\% uncertainty in the signal strength (ratio of observed to expected cross sections) for the SM Higgs boson is applied~\cite{HIG-13-004}.
Theoretical uncertainties arising from the underlying event and parton showering matching scale, PDF~\cite{Alekhin:2011sk} and the dependence on factorisation and normalisation scales are considered for signal. The PDF uncertainty is taken as the difference in the signal acceptance for the signal simulation with CTEQ6L1, MSTW2008NLO~\cite{MSTW2008NLO}, and NNPDF2.3NLO~\cite{NNPDF23NLO}  PDF sets, leading to a 10\% uncertainty.
A 20\% uncertainty in the signal normalisation  is applied to take into account the possible difference in the product of acceptance and efficiency between the LO sample generated with {\sc PYTHIA6.4} and the NLO sample generated by the {\sc MadGraph5\_aMC@NLO} generator~\cite{amcATNLO}.

The $\tauh$ and electron energy scales are among the systematic uncertainties affecting the $m_{\tau\tau}$ distributions.
To estimate the effects of these uncertainties, the electron  energy scale is changed by $1\%$ or by $2.5\%$ for electrons reconstructed in the barrel or in the endcap regions of the ECAL~\cite{EleMVA}, respectively, while the $\tauh$ energy scale is varied  by 3\%~\cite{Khachatryan:2015dfa}.
The top quark \pt reweighting correction, used for simulated \ttbar
events to match the observed \pt spectrum in a dedicated control region, is changed between
zero and twice the nominal value~\cite{Khachatryan:2015oqa,Chatrchyan:2012saa}.
The uncertainty in the $\tauh$ misidentification rate correction of the QCD multijet and $\PW$+jets background distributions has been taken into account.
To estimate this uncertainty, the $\tauh$ misidentification rate correction has been changed between
zero and twice its value.
An additional trigger uncertainty is applied to the \mtau~and \etau~final states to cover possible differences between collision data and simulated events
in the low-$\pt$  lepton region, where the trigger efficiency has not yet reached its plateau. These low-$\pt$ leptons are attributed an uncertainty that corresponds to half of the difference between the measured  and the plateau efficiencies.
Finally, uncertainties due to the limited number of simulated events, or the number of events in the control regions in data, are taken into account. These uncertainties are uncorrelated across the bins in each background distribution~\cite{BarlowBeeston}.

Among all systematic uncertainties, the ones that have the largest impact on the results are the $\tauh$ energy scale, the uncertainties related to the jet to muon, electron or $\tauh$ misidentification rates, and the uncertainties from the limited number of simulated events (or the observed events in data control regions). The impact of these individual uncertainties on the combined expected limit ranges between 5 and 10\% depending on $m_{\tau\tau}$.

\begin{table*}[tbp]
\begin{center}
\renewcommand{\arraystretch}{1.1}
\topcaption{Systematic uncertainties that affect the normalisation.}
\begin{tabular}{ll|rrr}
 \multicolumn{1}{c}{} & \multirow{2}{*}{Systematic source} & \multicolumn{3}{c}{Systematic uncertainty}\\
 \multicolumn{1}{c}{} & & \multicolumn{1}{c}{$\mu\tauh$} & \multicolumn{1}{c}{$\Pe\tauh$} & \multicolumn{1}{c}{$\Pe\mu$} \\
\cline{2-5}
\parbox[t]{3mm}{\multirow{17}{*}{\rotatebox[origin=c]{90}{Normalisation}}} & Integrated luminosity & 2.6\% & 2.6\% & 2.6\%    \\
& Muon ID/trigger & 2\% & \multicolumn{1}{c}{\NA} & 2\%  \\
& Electron ID/trigger & \multicolumn{1}{c}{\NA} & 2\% & 2\% \\
& $\tauh$ ID/trigger & 8\% & 8\% & \multicolumn{1}{c}{\NA} \\
& Muon to $\tauh$ misidentification rate & 30\% & \multicolumn{1}{c}{\NA} & \multicolumn{1}{c}{\NA} \\
& Electron to $\tauh$ misidentification rate & \multicolumn{1}{c}{\NA} & 30\% & \multicolumn{1}{c}{\NA} \\
& b tagging efficiency  & 1--4\% & 1--4\% & 1--4\%   \\
& b mistag rate  & 1--9\% & 1--9\% & 1--9\%   \\
& \MET scale & 1--2\% & 1--2\% & 1--2\%   \\
& $\PZ/\gamma^*\to \tau\tau$ normalisation& 3\% & 3\% & 3\% \\
& $\PZ/\gamma^*\to \tau\tau$ low-mass normalisation& 10\% & 10\% & 10\% \\
& QCD multijet normalisation & 20\% & 20\% & \multicolumn{1}{c}{\NA}  \\
& Reducible background normalisation & \multicolumn{1}{c}{\NA} & \multicolumn{1}{c}{\NA} & 30\%  \\
& $\PW$+jets normalisation & 30\% & 30\% & \multicolumn{1}{c}{\NA} \\
& \ttbar cross section & 10\% & 10\% & 10\%  \\
& Diboson cross section& 15\% & 15\% & 15\%  \\
& H$\to \tau \tau$ signal strength & 30\% & 30\% & 30\% \\
\cline{2-5}
\parbox[t]{3mm}{\multirow{4}{*}{\rotatebox[origin=c]{90}{Theory}}} & Underlying event and parton shower & 1--5\% & 1--5\% & 1--5\% \\
& Scales for A boson production & 10\% & 10\% & 10\% \\
& PDF for generating signal & 10\% & 10\% & 10\%\\
& NLO vs. LO & 20\% & 20\% & 20\%\\
\end{tabular}
\label{table:uncertainties}
\end{center}
\end{table*}

\section{Results}
\label{sec:Results}

\begin{figure*}[htbp]
\begin{center}
\includegraphics[scale=0.30]{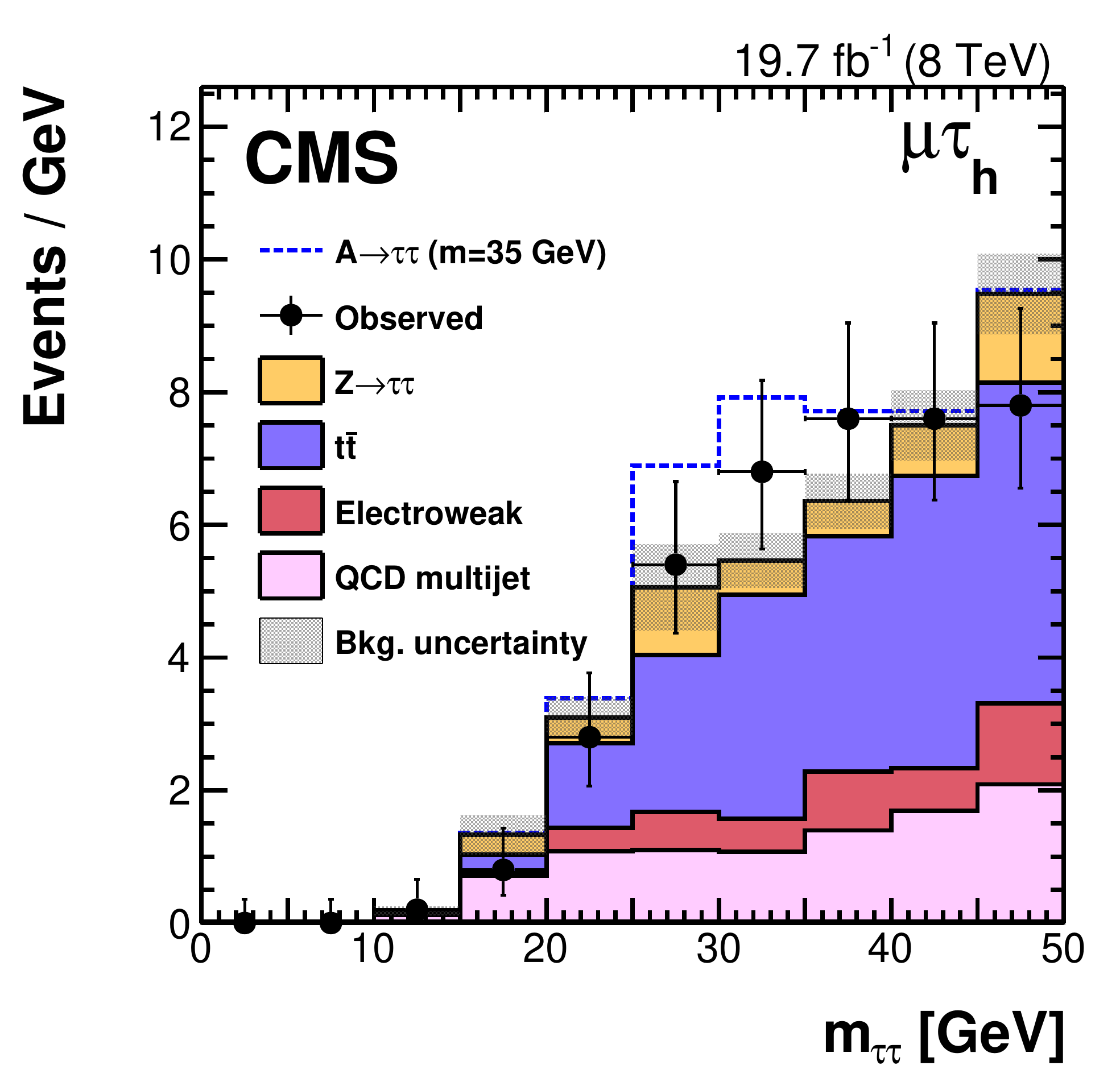}
\includegraphics[scale=0.30]{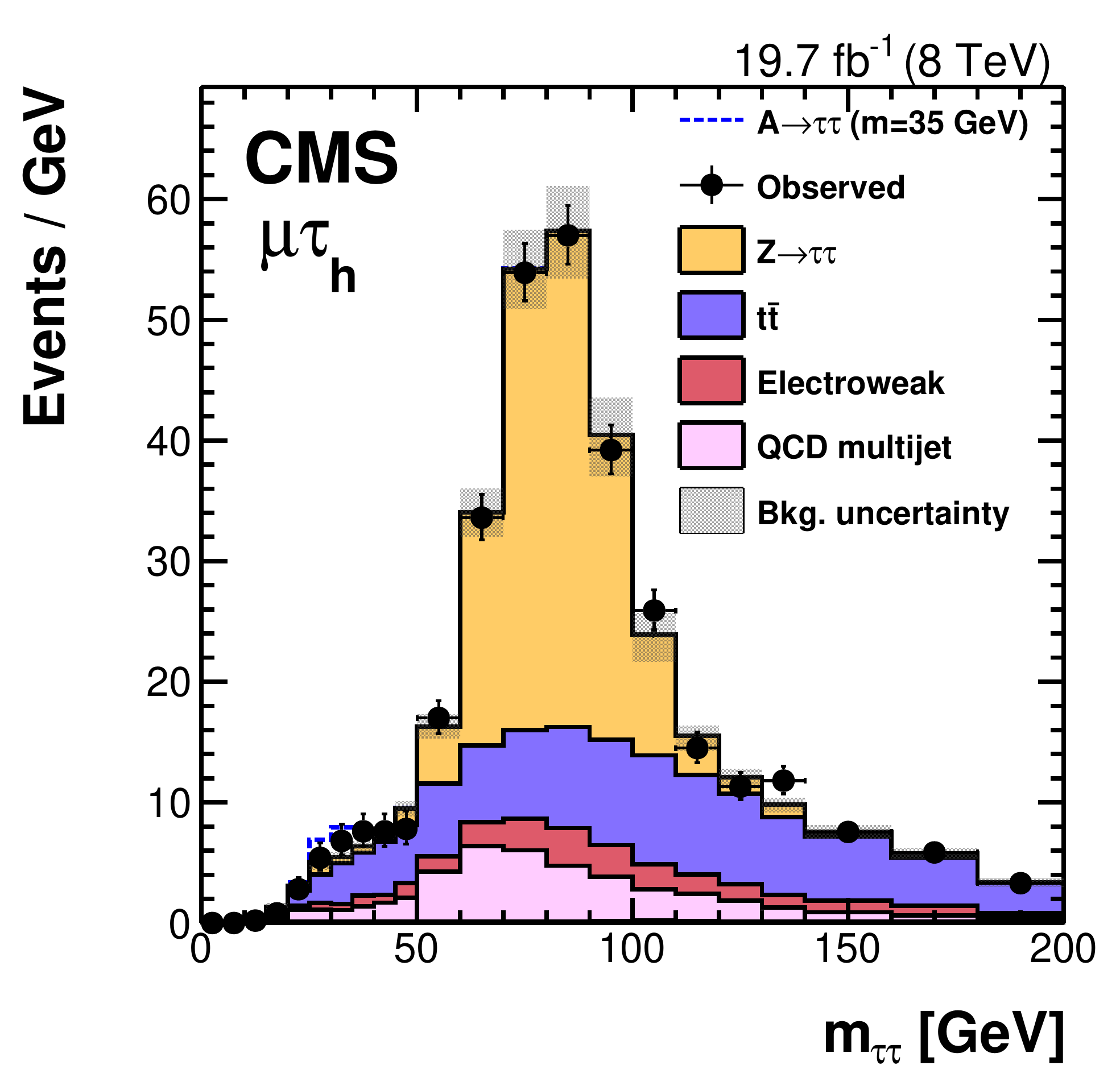}
\includegraphics[scale=0.30]{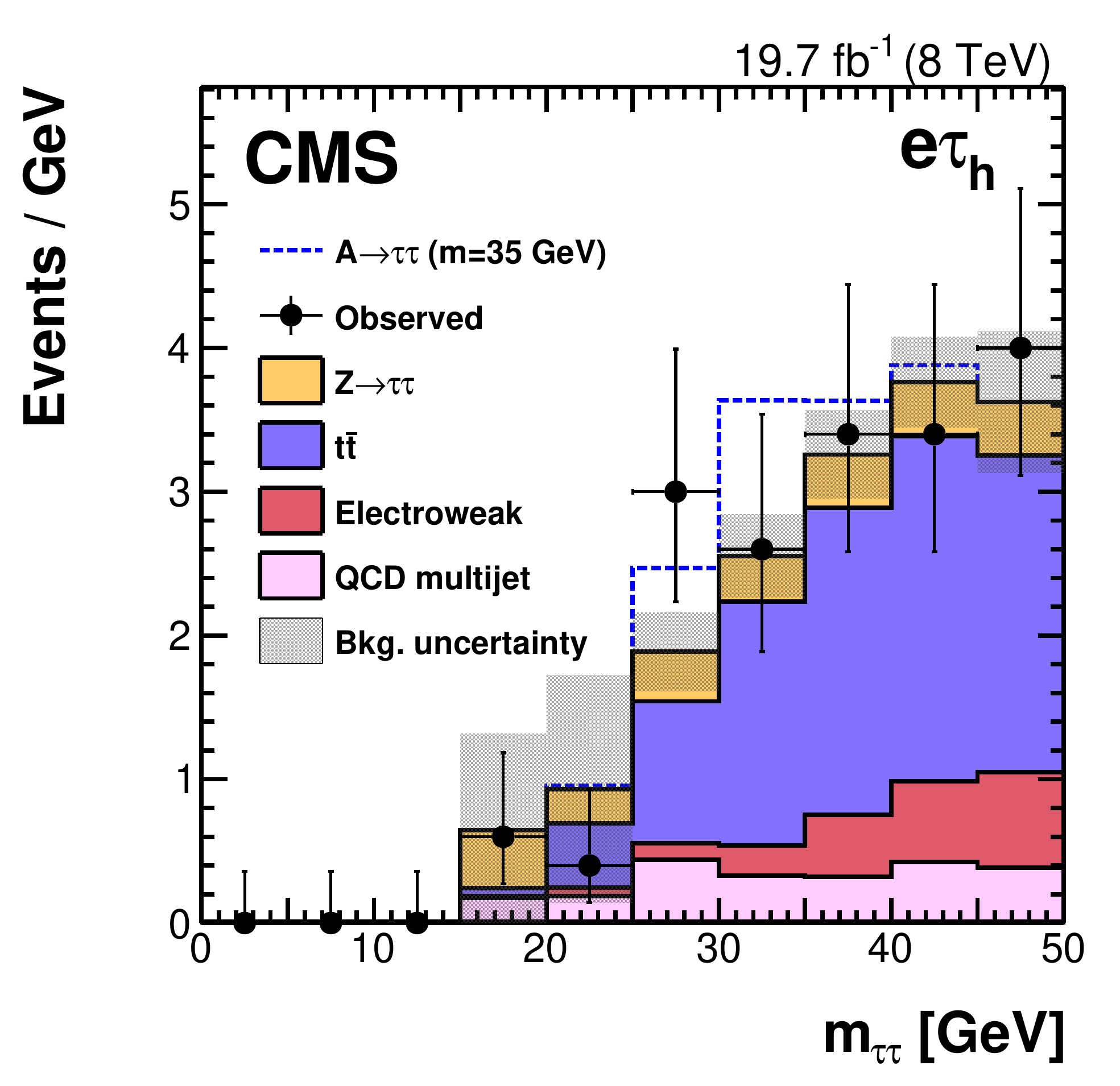}
\includegraphics[scale=0.30]{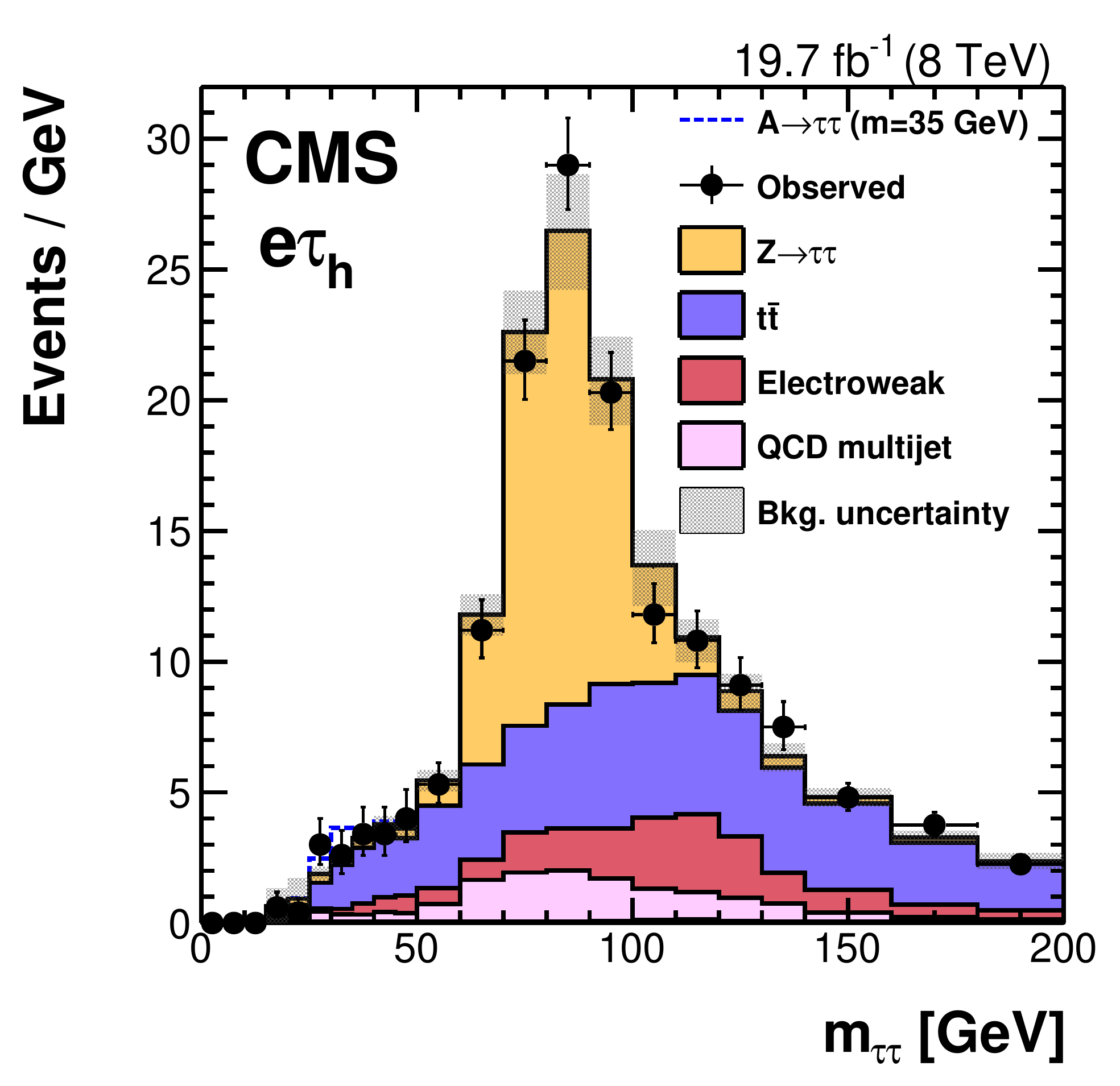}
\includegraphics[scale=0.30]{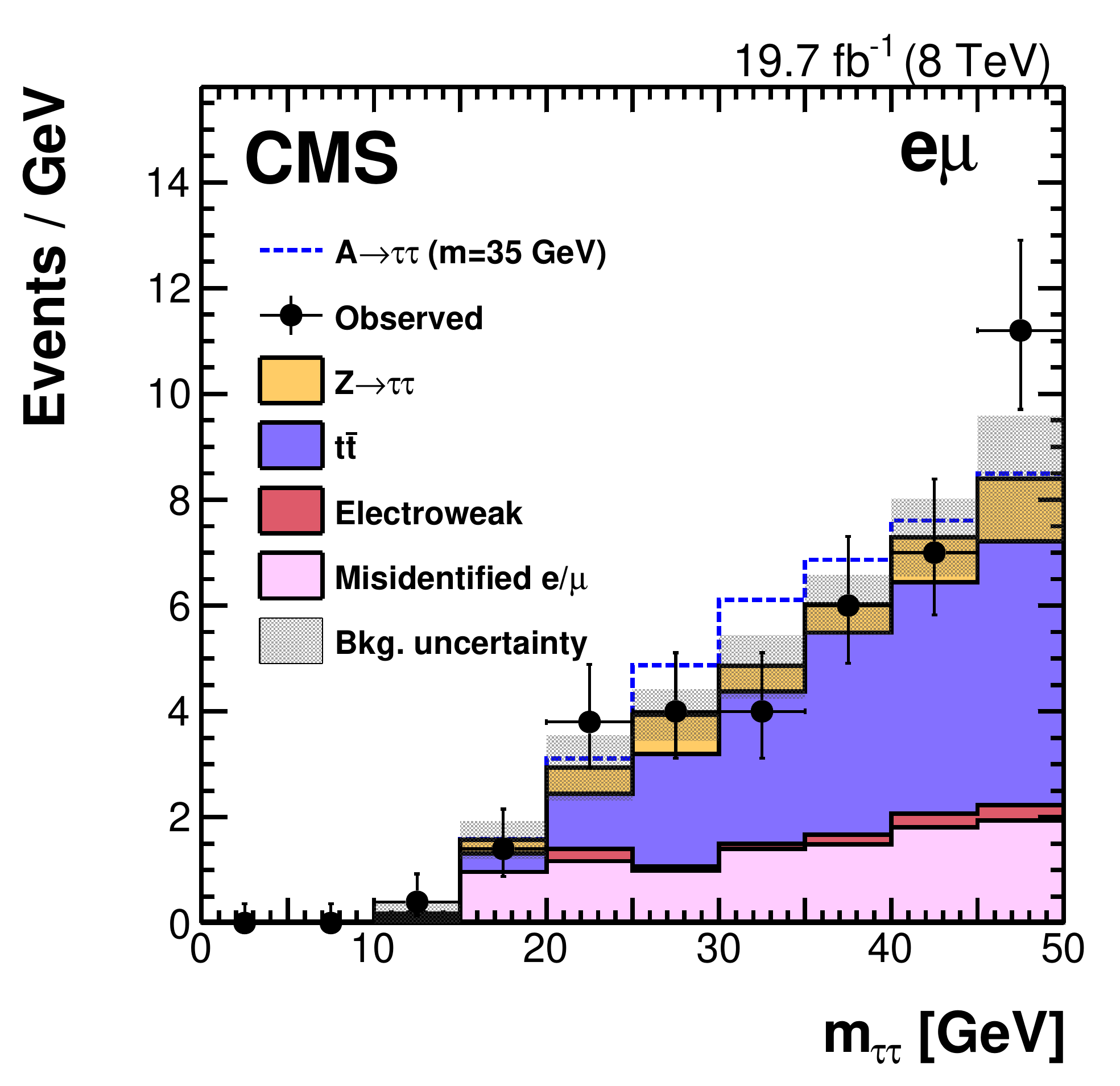}
\includegraphics[scale=0.30]{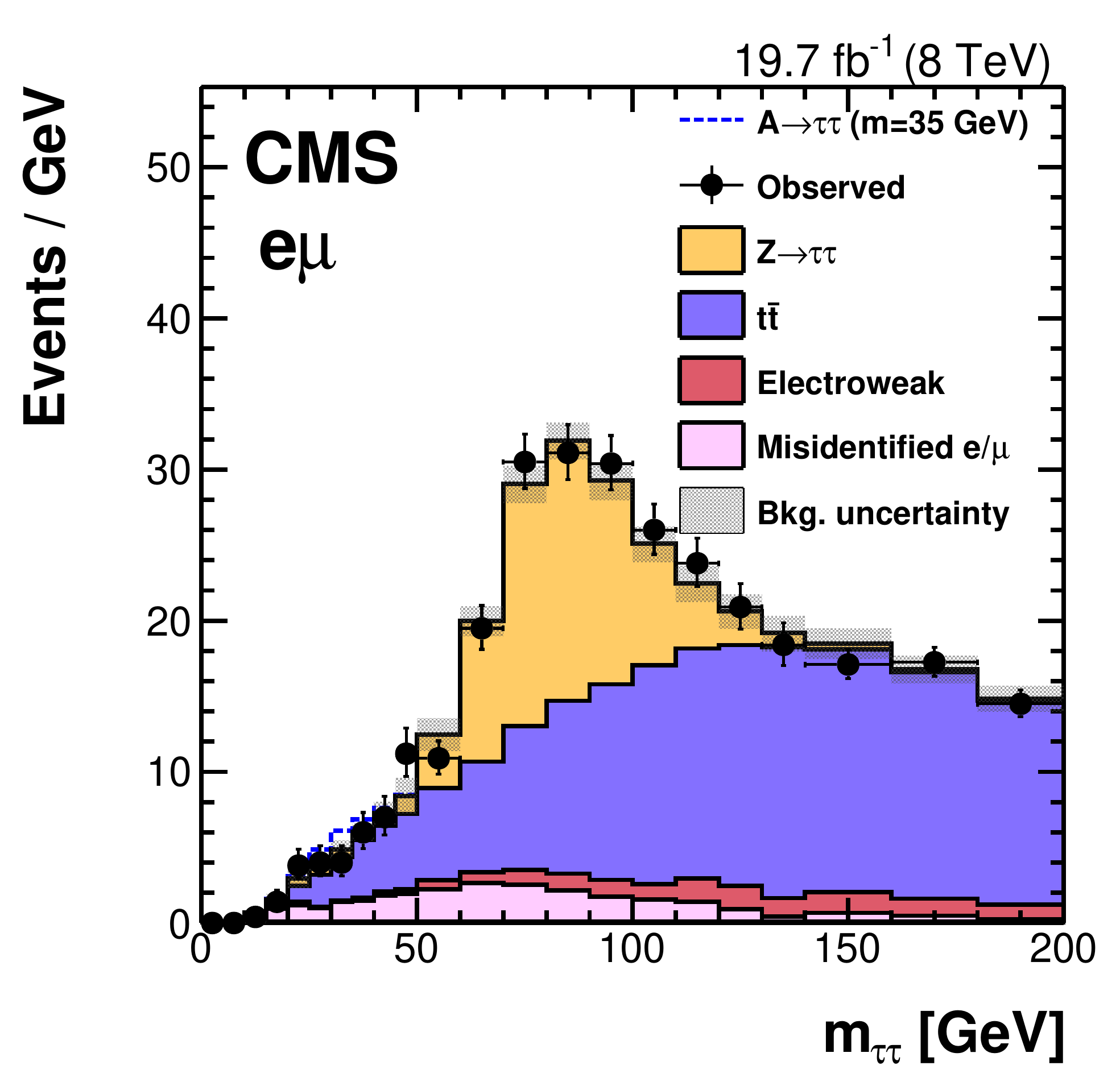}
\caption{Observed and predicted $m_{\tau\tau}$ distributions in the $\mu\tauh$ (top), $\Pe\tauh$ (middle), and $\Pe\mu$ (bottom) channels. The plots on the left are the zoomed-in versions for $m_{\tau\tau}$ distributions below 50\GeV.
A signal for a mass of $m_{\PSA}=35\GeV$ is shown for a cross section of 40\unit{pb}. In $\mu\tauh$ and $\Pe\tauh$ final states, the electroweak background is composed of $\PZ\to \Pe\Pe$, $\PZ \to \mu\mu$, $\PW$+jets, diboson, and single top quark contributions. In the $\Pe\mu$ final state, the electroweak background is composed of diboson and single top backgrounds, while the misidentified $\Pe/\mu$ background is due to QCD multijet and $\PW$+jets events. The contribution from the SM Higgs boson is negligible and therefore not shown. Expected background contributions are shown for the values of nuisance parameters (systematic uncertainties) obtained after fitting the signal + background hypothesis to the data.}
\label{fig:massplots}
\end{center}
\end{figure*}

The mass distributions for the $\mu\tauh$, e$\tauh$ and e$\mu$ channels are shown in Fig.~\ref{fig:massplots}. No significant excess of data is observed on top of the SM backgrounds.
A binned maximum likelihood fit has been applied simultaneously to all three distributions,
taking into account the systematic uncertainties as nuisance parameters. A log-normal probability distribution function is assumed for the nuisance parameters that affect the event yields of the various background contributions. Systematic uncertainties affecting the $m_{\tau\tau}$ distributions are assumed to have a Gaussian probability distribution function.

\begin{figure*}[htbp]
\begin{center}
\includegraphics[scale=0.26]{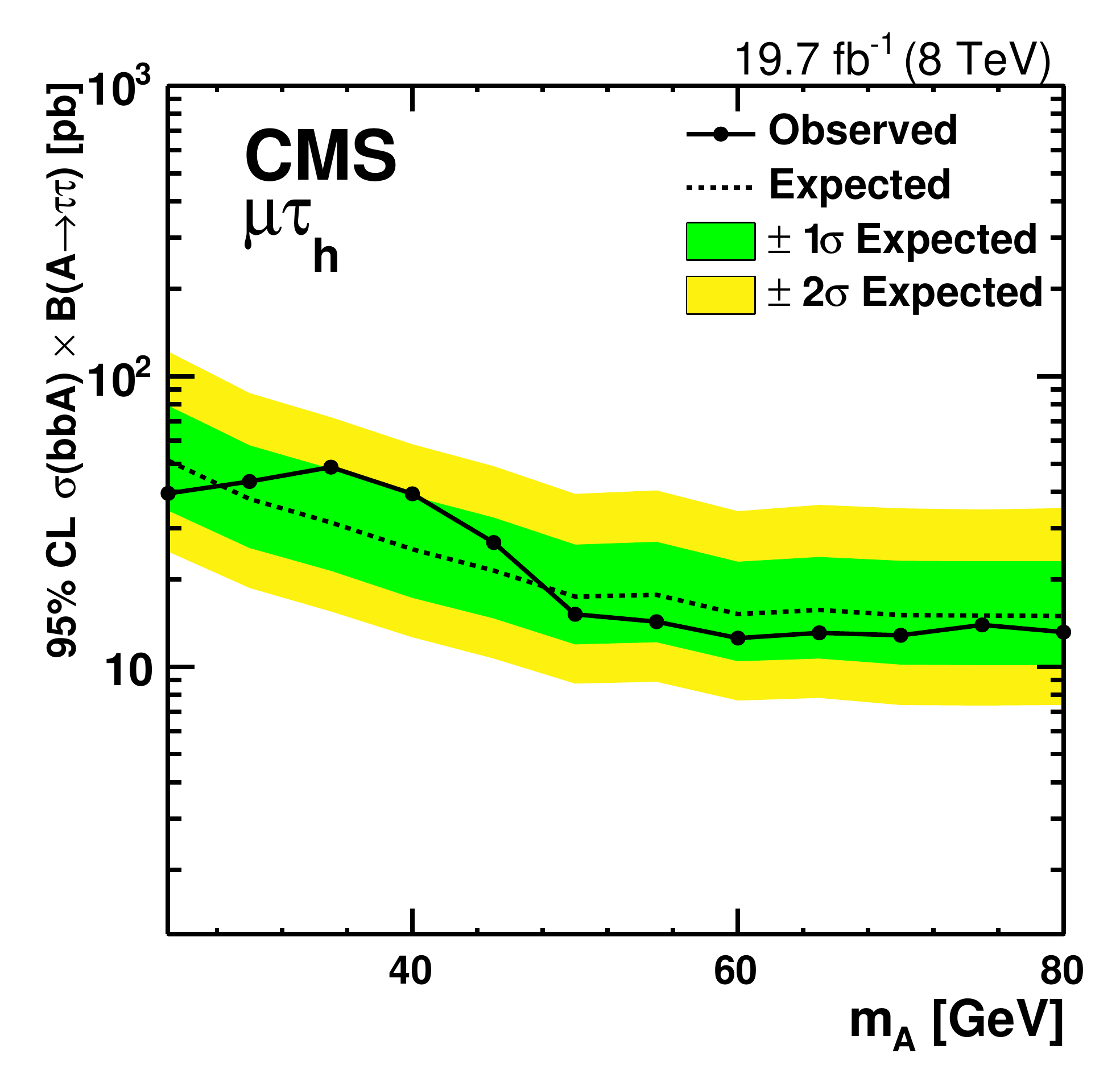}
\includegraphics[scale=0.26]{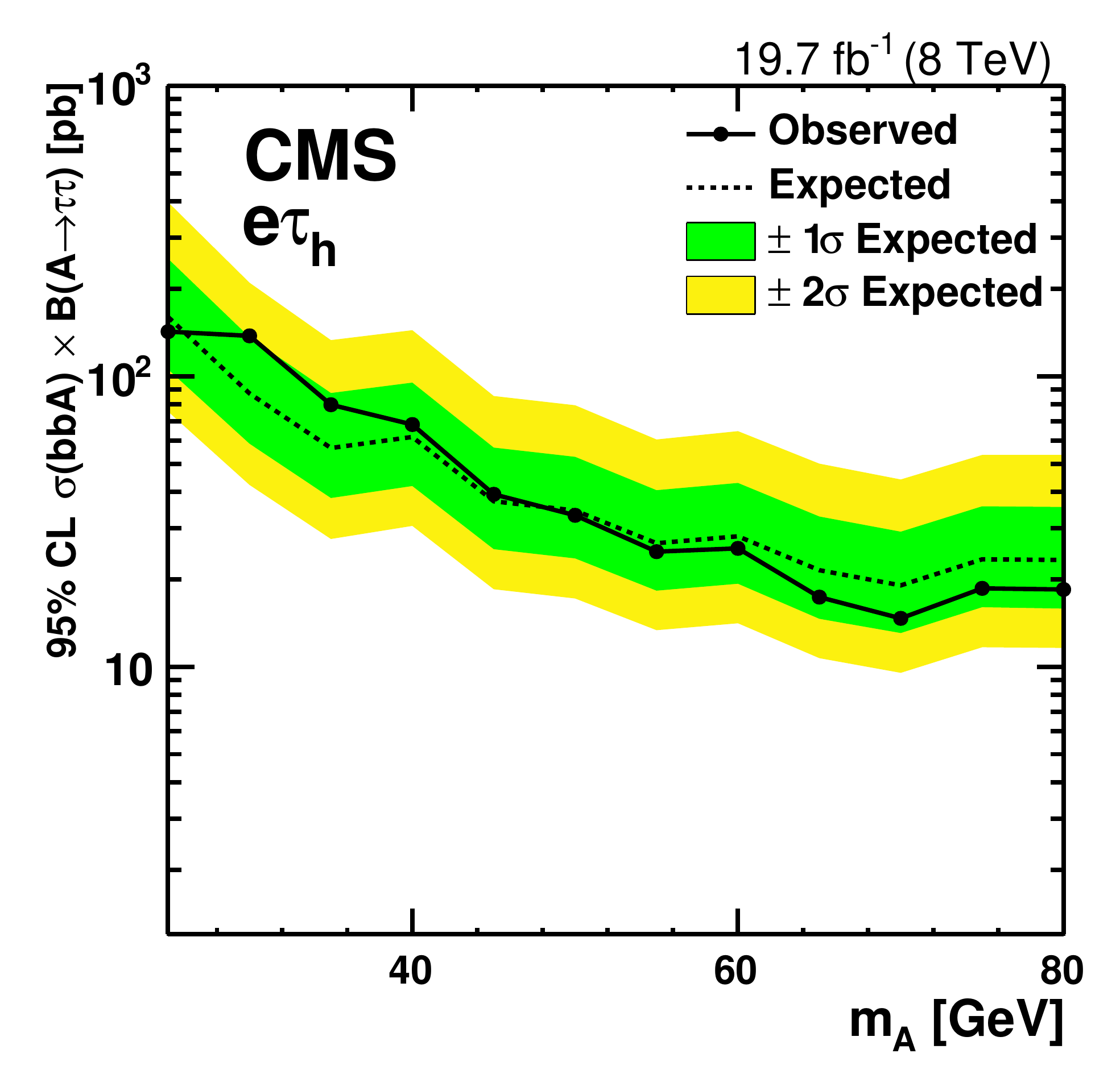}
\includegraphics[scale=0.26]{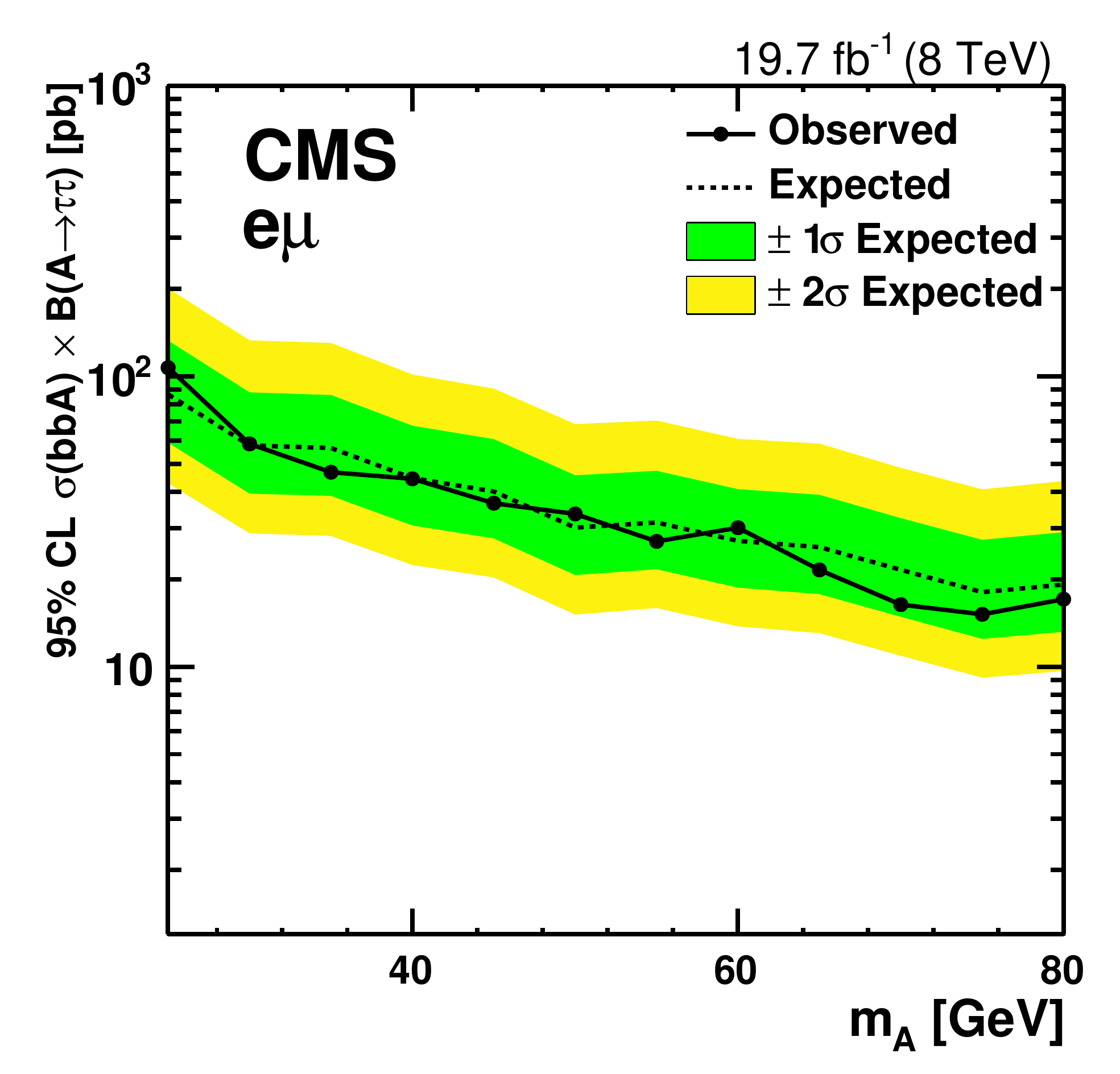}
\caption{Observed and expected upper limits at 95\% CL on the product of cross section and branching fraction for a light pseudoscalar Higgs boson produced in association with two b quarks, that decays to two $\tau$ leptons, in the $\mu\tauh$ (left), $\Pe\tauh$ (middle), and $\Pe\mu$ (right) channels. The 1$\sigma$ and 2$\sigma$ bands represent the 1 and 2 standard deviation uncertainties on the expected limits.}
\label{fig:individual_limits}
\end{center}
\end{figure*}

Upper limits on the product of cross section and branching fraction of the pseudoscalar Higgs boson to $\tau\tau$ are set at 95\% confidence level (CL) using the modified frequentist construction CL$_\mathrm{s}$~\cite{Read:2002hq,CLS2} and the procedure is described in Refs.~\cite{Cowan:2010js,LHC-HCG-Report}.
The observed and expected limits on the $\bbbar \PSA \to \bbbar \tau\tau$ process and the one and two standard deviation uncertainties on the expected limits are shown in Fig.~\ref{fig:individual_limits}.
Among the three channels, $\mu\tauh$ is the most sensitive one for the entire mass range because of the higher branching fraction relative to the $\Pe\mu$ channel, lower trigger and offline thresholds on the lepton $\pt$ relative to the $\Pe\tauh$ channel, and higher muon than electron identification efficiency.
Although background yields increase sharply with the mass, the acceptance of the signal grows faster, providing thereby more stringent limits
on the cross section at higher masses. The product of signal acceptance and efficiency in the $\mu\tauh$ channel changes from $1.5\times 10^{-5}$ at an A boson mass of 25\GeV to $6\times10^{-4}$ at $m_A=80$\GeV. In the $\Pe\tauh$ channel it ranges from $3\times 10^{-6}$ at 25\GeV to $2\times 10^{-4}$ at 80\GeV, and finally in the $\Pe\mu$ channel, it ranges from $1.3\times 10^{-5}$ at 25\GeV to $3.5\times 10^{-4}$ at 80\GeV. The trigger requirements and the \pt threshold of the leptons and $\tauh$s are the main factors in driving the signal acceptance and efficiency, especially at low masses.

\begin{figure}[htbp]
\begin{center}
\includegraphics[scale=0.35]{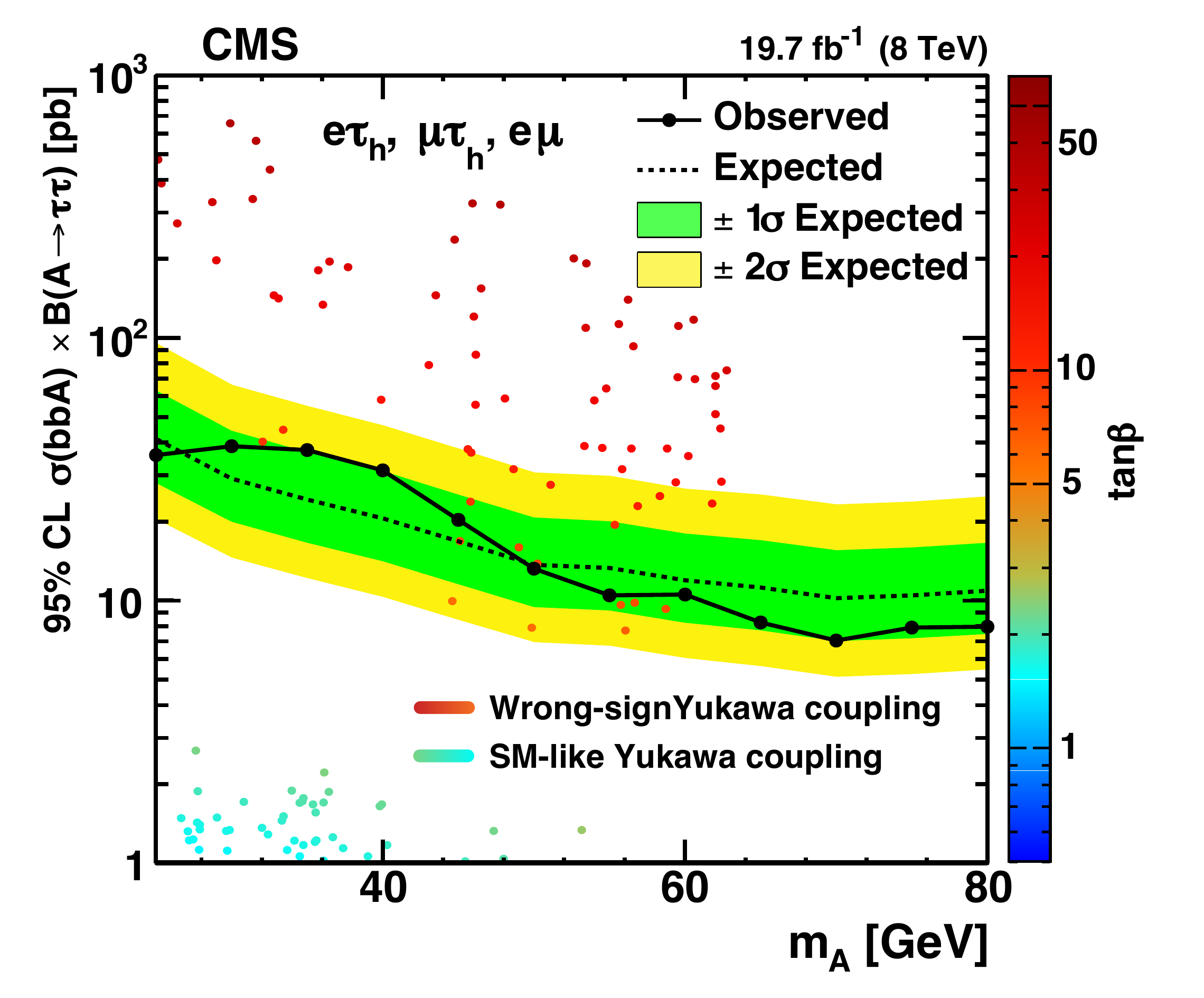}
\caption{Expected cross sections for Type~II 2HDM, superimposed on the expected and observed combined limits from this search. Cyan and green points, indicating small values of $\tan\beta$ as shown in the colour scale, have $\sin(\beta - \alpha) \approx 1$, $\cos(\beta - \alpha) > 0$, and low $m_{12}^{2}$, and correspond to models with SM-like Yukawa coupling, while red and orange points, with large $\tan\beta$, have $\sin(\beta + \alpha) \approx 1$, small cos$(\beta - \alpha) < 0$, and  $\tan\beta> 5$, and correspond to the models with a ``wrong sign" Yukawa coupling. Theoretically viable  points are shown only up  to $m_{\PSA} = m_{\Ph}/2$~\cite{JGunion8TeV}. (For interpretation of the references to colour in this figure legend, the reader is referred to the web version of this article.)}
\label{fig:cmb-theory}
\end{center}
\end{figure}

The upper limits from the combination of all final states are presented in Fig.~\ref{fig:cmb-theory}, with exact values quoted in Table~\ref{tab:cmb}. They range
from 7 to 39\unit{pb}  for A boson masses between 25 and 80\GeV.
In addition, superimposed in Fig.~\ref{fig:cmb-theory} are several typical production cross sections for the pseudoscalar Higgs boson produced in association with a pair of b quarks in Type~II 2HDM, for $m_{\PSA}$ less than half of the 125\GeV  Higgs boson ($\Ph$), and for $\mathcal{B} (\Ph \to \PSA\PSA) < 0.3$~\cite{JGunion8TeV}. The points are obtained from a series of scans in the 2HDM parameter space. Points with SM-like Yukawa coupling and small $\tan\beta$ have
$\sin(\beta - \alpha) \approx 1$, $\cos(\beta - \alpha) > 0$, and low $m_{12}^{2}$, while points with ``wrong sign" Yukawa coupling have $\sin(\beta \pm \alpha) \approx 1$, small $\cos(\beta - \alpha) < 0$, and $\tan\beta>5$.
While the combined results of the current analysis are not sensitive to the SM-like Yukawa coupling, they exclude the ``wrong sign" Yukawa coupling for almost the entire mass range, and more generally for $\tan\beta > 5$.
For masses greater than $m_{\Ph}/2$, where the constraint on $\mathcal{B} (\Ph \to \PSA\PSA ) < 0.3$ is automatically satisfied, the production cross section of the
pseudoscalar Higgs boson in association with a pair of b quarks is much larger~\cite{JGunion13TeV}; consequently, the exclusion limit extends to masses up to 80\GeV.

\begin{table*}[!htbp]
\newcolumntype{.}{D{.}{.}{-1}}
\renewcommand{\arraystretch}{1.05}
\begin{center}
\topcaption{Expected and observed combined upper limits at 95\% CL in pb, along with their 1 and 2
standard deviation uncertainties, in the product of cross section and branching fraction for
pseudoscalar Higgs bosons produced in association with \bbbar pairs.}
\begin{tabular}{c|.....|.}
     \multirow{2}{*}{$m_{\PSA}(\GeVns{})$} &     \multicolumn{5}{c|}{Expected limit (pb)} & \multicolumn{1}{c}{\multirow{2}{*}{Observed limit (pb)}} \\
          &      \multicolumn{1}{c}{$-2\sigma$} &       \multicolumn{1}{c}{$-1\sigma$} &           \multicolumn{1}{c}{Median} &       \multicolumn{1}{c}{$+1\sigma$} &       \multicolumn{1}{c|}{$+2\sigma$} &      \\
          \hline
                25 &               20.4 &               28.1 &               41.3 &               63.1 &               95.5 &               35.8  \\
                30 &               14.6 &               20.0 &               29.1 &               44.3 &               66.3 &               38.7  \\
                35 &               12.2 &               16.6 &               24.3 &               36.7 &               55.1 &               37.4  \\
                40 &               10.3 &               14.1 &               20.6 &               31.1 &               46.5 &               31.3  \\
                45 &               8.4 &               11.6 &               16.8 &               25.3 &               37.9 &               20.3  \\
                50 &               7.0 &               9.5 &               13.7 &               20.7 &               30.8 &               13.2  \\
                55 &               6.7 &               9.2 &               13.3 &               20.1 &               29.9 &               10.5  \\
                60 &               6.1 &               8.2 &               12.0 &               18.0 &               26.7 &               10.6  \\
                65 &               5.6 &               7.7 &               11.2 &               17.0 &               25.4 &               8.3  \\
                70 &               5.1 &               7.0 &               10.2 &               15.6 &               23.3 &               7.1  \\
                75 &               5.3 &               7.2 &               10.5 &               15.9 &               23.8 &               7.9  \\
                80 &               5.5 &               7.5 &               10.9 &               16.6 &               25.0 &               8.0  \\

\end{tabular}
\label{tab:cmb}
\end{center}
\end{table*}

\section{Summary}
\label{Summary}

A search by the CMS experiment for a light pseudoscalar Higgs boson produced in association with a \bbbar pair and decaying to a pair of $\tau$ leptons is reported. Three final states: \mtau, \etau, and \emu, are used where $\tauh$ represents a hadronic $\tau$ decay.
The results are based on proton-proton collision data accumulated at a centre-of-mass energy of 8\TeV, corresponding to an integrated luminosity
of 19.7\fbinv. Pseudoscalar boson masses between 25 and 80\GeV are probed.  No evidence for a pseudoscalar boson is found and upper limits are set on the product of cross section and branching fraction to $\tau$ pairs between 7 and 39\unit{pb} at the 95\% confidence level.
This excludes pseudoscalar A bosons with masses between 25 and 80\GeV, with SM-like Higgs boson negative couplings to down-type fermion, produced in association with \bbbar pairs, in Type II, two-Higgs-doublet models.

\begin{acknowledgments}
We congratulate our colleagues in the CERN accelerator departments for the excellent performance of the LHC and thank the technical and administrative staffs at CERN and at other CMS institutes for their contributions to the success of the CMS effort. In addition, we gratefully acknowledge the computing centres and personnel of the Worldwide LHC Computing Grid for delivering so effectively the computing infrastructure essential to our analyses. Finally, we acknowledge the enduring support for the construction and operation of the LHC and the CMS detector provided by the following funding agencies: BMWFW and FWF (Austria); FNRS and FWO (Belgium); CNPq, CAPES, FAPERJ, and FAPESP (Brazil); MES (Bulgaria); CERN; CAS, MoST, and NSFC (China); COLCIENCIAS (Colombia); MSES and CSF (Croatia); RPF (Cyprus); MoER, ERC IUT and ERDF (Estonia); Academy of Finland, MEC, and HIP (Finland); CEA and CNRS/IN2P3 (France); BMBF, DFG, and HGF (Germany); GSRT (Greece); OTKA and NIH (Hungary); DAE and DST (India); IPM (Iran); SFI (Ireland); INFN (Italy); MSIP and NRF (Republic of Korea); LAS (Lithuania); MOE and UM (Malaysia); CINVESTAV, CONACYT, SEP, and UASLP-FAI (Mexico); MBIE (New Zealand); PAEC (Pakistan); MSHE and NSC (Poland); FCT (Portugal); JINR (Dubna); MON, RosAtom, RAS and RFBR (Russia); MESTD (Serbia); SEIDI and CPAN (Spain); Swiss Funding Agencies (Switzerland); MST (Taipei); ThEPCenter, IPST, STAR and NSTDA (Thailand); TUBITAK and TAEK (Turkey); NASU and SFFR (Ukraine); STFC (United Kingdom); DOE and NSF (USA).

Individuals have received support from the Marie-Curie programme and the European Research Council and EPLANET (European Union); the Leventis Foundation; the A. P. Sloan Foundation; the Alexander von Humboldt Foundation; the Belgian Federal Science Policy Office; the Fonds pour la Formation \`a la Recherche dans l'Industrie et dans l'Agriculture (FRIA-Belgium); the Agentschap voor Innovatie door Wetenschap en Technologie (IWT-Belgium); the Ministry of Education, Youth and Sports (MEYS) of the Czech Republic; the Council of Science and Industrial Research, India; the HOMING PLUS programme of the Foundation for Polish Science, cofinanced from European Union, Regional Development Fund; the OPUS programme of the National Science Center (Poland); the Compagnia di San Paolo (Torino); MIUR project 20108T4XTM (Italy); the Thalis and Aristeia programmes cofinanced by EU-ESF and the Greek NSRF; the National Priorities Research Program by Qatar National Research Fund; the Rachadapisek Sompot Fund for Postdoctoral Fellowship, Chulalongkorn University (Thailand); and the Welch Foundation, contract C-1845.
\end{acknowledgments}

\bibliography{auto_generated}
\cleardoublepage \appendix\section{The CMS Collaboration \label{app:collab}}\begin{sloppypar}\hyphenpenalty=5000\widowpenalty=500\clubpenalty=5000\input{HIG-14-033-authorlist.tex}\end{sloppypar}
\end{document}

%% file: HIG-14-033-authorlist.tex
\textbf{Yerevan Physics Institute,  Yerevan,  Armenia}\\*[0pt]
V.~Khachatryan, A.M.~Sirunyan, A.~Tumasyan
\vskip\cmsinstskip
\textbf{Institut f\"{u}r Hochenergiephysik der OeAW,  Wien,  Austria}\\*[0pt]
W.~Adam, E.~Asilar, T.~Bergauer, J.~Brandstetter, E.~Brondolin, M.~Dragicevic, J.~Er\"{o}, M.~Flechl, M.~Friedl, R.~Fr\"{u}hwirth\cmsAuthorMark{1}, V.M.~Ghete, C.~Hartl, N.~H\"{o}rmann, J.~Hrubec, M.~Jeitler\cmsAuthorMark{1}, V.~Kn\"{u}nz, A.~K\"{o}nig, M.~Krammer\cmsAuthorMark{1}, I.~Kr\"{a}tschmer, D.~Liko, T.~Matsushita, I.~Mikulec, D.~Rabady\cmsAuthorMark{2}, B.~Rahbaran, H.~Rohringer, J.~Schieck\cmsAuthorMark{1}, R.~Sch\"{o}fbeck, J.~Strauss, W.~Treberer-Treberspurg, W.~Waltenberger, C.-E.~Wulz\cmsAuthorMark{1}
% ...
\vskip\cmsinstskip
% ...
1:~~Also at Vienna University of Technology, Vienna, Austria\\
2:~~Also at CERN, European Organization for Nuclear Research, Geneva, Switzerland\\
% ...
